\documentclass[
aps,%
12pt,%
final,%
notitlepage,%
oneside,%
onecolumn,%
nobibnotes,%
nofootinbib,% 
superscriptaddress,%
noshowpacs,%
centertags]%
{revtex4}

\usepackage{bm}
\usepackage{graphics}
\usepackage{rotating}
\usepackage{amsmath}
\usepackage{amssymb}
\usepackage{epsfig}

\begin{document}
%\selectlanguage{english}

\title{PRODUCTION OF HEAVY QUARK BOUND STATES IN RARE EXCLUSIVE DECAYS OF HIGGS BOSON}

\author{\firstname{F.~A.} \surname{Martynenko}\footnote
{Talk presented at the session-conference "Physics of Fundamental Interactions" 
dedicated to the 70th anniversary of the birth of RAS Academician Valery Anatolyevich Rubakov 
17-21 February 2025, https://indico.inr.ac.ru/event/5/}
\affiliation{Samara University, Samara, Russia}}

\author{\firstname{A.~P.} \surname{Martynenko}}
\affiliation{Samara University, Samara, Russia}

\author{\firstname{A.~V.} \surname{Eskin}}
\affiliation{Samara University, Samara, Russia}

\begin{abstract}
The processes of pair production
of S- and P- wave charmonium, as well as the production of the tetraquark
$(cc\bar c\bar c)$ in the Higgs boson decays are investigated within the framework 
of relativistic quark model.
Relativistic decay amplitudes
are constructed and decay widths are calculated taking into account relativistic corrections 
both in the decay amplitude and in the wave function of bound heavy quarks.
\end{abstract}

\maketitle

\section{Introduction}

After the discovery of the Higgs boson \cite{atlas,cms}, the next level of studying electroweak
physics began to form, connected with precision measurements of the processes 
of its production and decay.
Among the rare decays of the Higgs boson, one can distinguish processes in which bound 
states of heavy quarks are formed. On the one hand, such reactions allow
to determine the values of various constants of particle interaction in the Higgs sector.
On the other hand, an opportunity opens up for additional check of the mechanisms of
the production of quark bound states.

The reactions of charmonium and bottomonium production in the Higgs boson decays have been 
studied for a long time both within the nonrelativistic
QCD and within the quark models \cite{luchinsky,bodwin,enterria} (see references to other works in \cite{enterria}). 
These studies show that exclusive decays of the Higgs boson with the production of heavy quarkonia 
can provide information about the coupling constants of quarks of different generations and 
serve to search for New Physics beyond the Standard Model. Pair production of heavy quarkonia 
in the Higgs boson decays is a particular class of rare processes that have also begun 
to be studied experimentally \cite{Tumasyan_2023,pdg}. Previous studies focused mainly 
on the production of $S$-states charmonium \cite{gao,faustov2023relativistic}. In this work 
we have expanded the scope of the study to include also the production of $P$ - wave states, 
as in the case of other reactions \cite{faustov2022higgs,em}.

Our studies of quarkonia production processes in the Higgs boson decays are carried out within 
the relativistic quasipotential approach, which is based on the microscopic picture 
of the interaction of quarks and gauge bosons and is presented in detail in our previous works \cite{faustov2022higgs,faustov2023relativistic,martynenko2024production}. This approach allows 
us to consistently take into account relativistic effects in decay processes, 
which are necessary for their reliable description.

\section{Pair production of $S$- and $P$-wave charmonia}

\begin{figure}[htbp]
\centering
\includegraphics[scale=0.8]{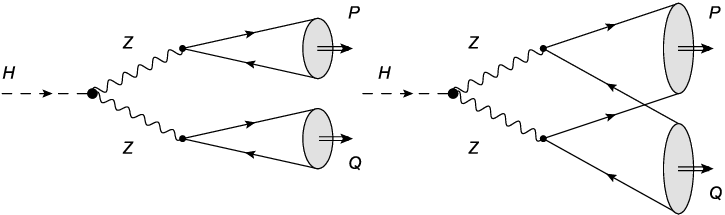}
\includegraphics[scale=1.0]{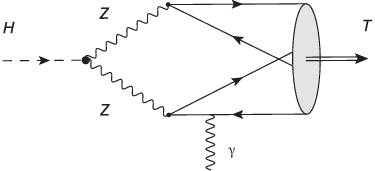}
\caption{$ZZ$ boson mechanism of pair charmonium production and tetraquak. 
$P$, $Q$ are four-momenta of quarkonia, $T$ is four-momentum of tetraquak.}
\label{zz_mech}
\end{figure}

\begin{figure}[htbp]
\centering
\includegraphics[width=0.6\textwidth]{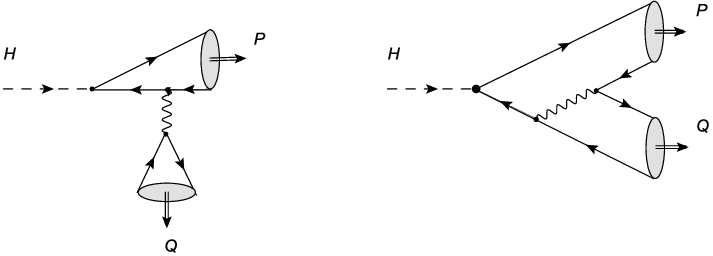}
\caption{Quark - photon mechanism of charmonium pair production. $P$, $Q$ are four-momenta of quarkonia.}
\label{qph_mech}
\end{figure}

There are various possible mechanisms for the formation of pairs of bound quark states in the Higgs 
boson decay. We consider the main mechanisms shown in Fig.~\ref{zz_mech}, \ref{qph_mech}, \ref{qgl_mech}, 
which, as shown by previous studies, make the main contribution to the decay width. Let us consider 
the general formalism for calculating the decay width using the example of the $ZZ$-boson mechanism, 
the amplitude of which is shown in Fig.~\ref{zz_mech}. The first amplitude gives direct charmonium 
production, in which each of the two $Z$-bosons
transitions to the final state $(c\bar c)$. The second amplitude describes the cross production mechanism, 
which is characterized by the formation of quark-antiquark pairs from
different $Z$-bosons.

For the $ZZ$-boson mechanism there are two vertices of particle interaction
$H\to ZZ$ and $Z\to c\bar c$ \cite{CORE}:
\begin{equation}
\label{f1}
V^{\alpha\beta}=\frac{2e}{\sin 2\theta_W} M_Z g^{\alpha\beta},~~~
\Gamma^{\alpha}=\frac{e}{\sin 2\theta_W} \gamma^\alpha
\left[\frac{1}{2}(1-\gamma_5)-a_c\right],~~~a_c=2Q_c\sin^2\theta_W,
\end{equation}
where $M_Z$ is the Z-boson mass, $Q_c$ is the $c$ quark charge, $\theta_W$ is the 
Weinberg angle.

In the considered pair production processes, one of the charmoniums is produced in the $S$-state, 
and the second one in the $P$-state. There are four $(c\bar c)$ states with orbital angular momentum 
$L=1$, which are presented in the Table~\ref{tab_mes_const}. The direct and cross amplitudes 
for the production of pseudoscalar $({\cal P})$ and vector $({\cal V})$ mesons have the form:
\begin{equation}
\label{f3}
{\cal M}_{ZZ}^{(1)}=\frac{2e^3 g^{\lambda\sigma}}{\sin^32\theta_W}
\int\frac{d{\bf p}}{(2\pi)^3}Tr\Bigl\{\Psi_{\cal V}(p,P)\Gamma^\alpha\Bigr\}
\int\frac{d{\bf q}}{(2\pi)^3}Tr\Bigl\{\Psi_{\cal P}(q,Q)\Gamma^\beta\Bigr\}
D^{\lambda\alpha}(P)D^{\sigma\beta}(Q),
\end{equation}
\begin{equation}
\label{f4}
{\cal M}_{ZZ}^{(2)}=\frac{2e^3 g^{\lambda\sigma}}{\sin^32\theta_W}
\int\frac{d{\bf p}}{(2\pi)^3}\int\frac{d{\bf q}}{(2\pi)^3}Tr\Bigl\{
\Psi_{\cal V}(p,P)\Gamma^\alpha\Psi_{\cal P}(q,Q)\Gamma^\beta\Bigr\}
D^{\lambda\alpha}(P+Q)D^{\sigma\beta}(P+Q).
\end{equation}

The initial expression for the amplitudes \eqref{f3}-\eqref{f4} is a convolution of the amplitudes for the production 
of free quarks with the Bethe-Salpeter wave functions describing the mesons. After a series of transformations, 
they take the form \eqref{f3}-\eqref{f4} \cite{em}, where
relativistic wave functions of pseudoscalar $\Psi_{\cal P}$ and vector 
$\Psi_{\cal V}$ bound states of quarks have the form:
\begin{eqnarray}
\label{f5}
\Psi_{\cal P}(p,P)&=&\frac{\Psi({\bf p})}{[\frac{\epsilon(p)}{m}\frac{(\epsilon(p)+m)}{2m}]}
\left[\frac{\hat v_1-1}{2}+\hat
v_1\frac{{\bf p}^2}{2m(\epsilon(p)+ m)}-\frac{\hat{p}}{2m}\right] \times \cr
&&\gamma_5(1+\hat v_1) \left[\frac{\hat
v_1+1}{2}+\hat v_1\frac{{\bf p}^2}{2m(\epsilon(p)+m)}+\frac{\hat{p}}{2m}\right],
\end{eqnarray}
\begin{eqnarray}
\label{f6}
\Psi_{\cal V}(q,Q)&=&\frac{\Psi({\bf q})}
{[\frac{\epsilon(q)}{m}\frac{(\epsilon(q)+m)}{2m}}
\left[\frac{\hat v_2-1}{2}+\hat v_2\frac{{\bf q}^2}{2m(\epsilon(q)+
m)}+\frac{\hat{q}}{2m}\right] \times \cr 
&&\hat{\varepsilon}(Q,S_z)(1+\hat v_2) 
\left[\frac{\hat v_2+1}{2}+\hat v_2\frac{{\bf q}^2}{2m(\epsilon(q)+ m)}-\frac{\hat{q}}{2m}\right],
\end{eqnarray}
$v_1=P/M_1$, $v_2=Q/M_2$, $M_{1,2}$ are the meson masses, $m$ is the $c$-quark mass,
$\varepsilon^\lambda(Q,S_z)$ is the charmonium spin four-vector,
$\epsilon(p)=\sqrt{p^2+m^2}$ is relativistic energy of quarks.
$\Psi({\bf p})$ is the charmonium wave function in the rest frame.
The momenta of quarks and antiquarks $p_{1,2}$, $q_{1,2}$ can be conveniently expressed 
through total and relative momenta as:
\begin{equation}
\label{f6a}
p_{1,2} = \frac{1}{2}P \pm p,~(p \cdot P) = 0,~q_{1,2} = \frac{1}{2}Q \pm q,~(q \cdot Q) = 0.
\end{equation}

The wave functions \eqref{f5}, \eqref{f6} are obtained as a result of the Lorentz 
transformation from the rest frame to the moving reference frame with four-momenta $P$, $Q$ 
\cite{faustov2022higgs}. Calculating the traces in \eqref{f3}-\eqref{f4}, we introduce the polarization 
vector of the orbital motion of $P$-wave charmonium:
\begin{equation}
\label{f7}
\int\frac{d{\bf q}}{(2\pi)^3}q_\mu \psi_{LL_z}({\bf q})=-i\varepsilon_\mu(L_Z)
\sqrt{\frac{3}{4\pi}}R'_{\cal P}(0),
\end{equation}
where $R'_P(0)$ is the derivative of radial wave function at zero. When adding the spin and orbital 
angular momenta, we obtain states with total momentum $J=0,~1,~2$:
\begin{equation}
\label{f8}
\sum_{L_z,S_z}<1,L_z;1,S_z|J,J_z>\varepsilon_\alpha(v_2,L_z)
\varepsilon_\beta(v_2,S_z)=
\begin{cases}
\frac{1}{\sqrt{3}}(g_{\alpha\beta}-v_{2\alpha}v_{2\beta}),~J=0,\\
\frac{i}{\sqrt{2}}\varepsilon_{\alpha\beta\mu\nu}v_2^\mu\varepsilon^\nu(v_2,J_z),~J=1,\\
\varepsilon_{\alpha\beta}(v_2,J_z),~J=2.\\
\end{cases}
\end{equation}

In the denominators of the $Z$-boson propagators, we can neglect relative quark momenta $p$, $q$ 
compared to the $Z$-boson mass. In the numerator of the amplitudes, we take into account the momenta 
of relative motion, extracting corrections of order of 
$O({\bf p}^2/m^2)$, $O({\bf q}^2/m^2)$.
Choosing the $Z$-boson propagator in the unitary gauge, we represent total amplitude 
${\cal M}_{ZZ}^{(1)} +
{\cal M}_{ZZ}^{(2)}$ in the case of the $H\to \eta_c+\chi_{c1}$ decay as follows:
\begin{equation}
\label{f10}
{\cal M}(H\to \eta_c+\chi_{c1}) = 
\frac{\sqrt{3}}{4\sqrt{2}\pi} \tilde R_{\cal S}(0)\tilde R'_{\cal P}(0) 
\frac{2e^3}{\sin^3\theta_W} \frac{(1-\omega_{1}-\tilde\omega_{1}+\omega_{1}\tilde\omega_{1})}
{mM_Z^3} \left( v_1\varepsilon(v_2,J_z)\right) \times 
\end{equation}
\begin{displaymath}
\Bigl[
\frac{1}{(1-r_4^2)}-
\frac{ \left( (r_3^2-r_4^2)[r_3r_4(r_1^2+r_2^2-1)-r_1r_2(r_3^2+r_4^2-8)]+32r_1r_2 \right)
\bigl(\frac{1}{2}-a_c+a_c^2\bigr)}{(r_5^2-4)^2r_1r_2}
\Bigl],
\end{displaymath}
where parameters are introduced that are determined by the ratios of the particle masses:
\begin{equation}
\label{f12}
r_1=\frac{M_1}{M_H},~~~r_2=\frac{M_2}{M_H},~~~r_3=\frac{M_1}{M_Z},~~~r_4=\frac{M_2}{M_Z},~~~
r_5=\frac{M_H}{M_Z}.
\end{equation}

Relativistic parameters $\omega_1$, $\tilde\omega_1$ are defined below.
In the case of the decay $H\to J/\Psi+h_{c}$ there is only cross amplitude contribution
which can be transformed to the form:
\begin{equation}
\label{f13}
{\cal M}(H\to J/\Psi+h_{c}) = 
\frac{\sqrt{3}}{4\sqrt{2}\pi} \tilde R_{\cal S}(0)\tilde R'_{\cal P}(0)
\frac{e^3}{2\sin^3\theta_W mM_Z^3}\varepsilon_{\mu\nu\lambda\sigma}
v_1^\mu v_2^\nu \varepsilon^\lambda(v_2,J_z)\varepsilon^\sigma(v_1,S_z)
\times
\end{equation}
\begin{displaymath}
(1-\omega_{1}-\tilde\omega_{1}+\omega_{1}\tilde\omega_{1})a_c(1-a_c)
\frac{\left( (r_3^2-r_4^2)[r_3r_4(1-r_1^2-r_2^2)+r_1r_2(r_3^2+r_4^2-8)] \right)}{(r_5^2-4)^2r_1r_2}.
\end{displaymath}

The expressions for amplitudes \eqref{f10}, \eqref{f13} contain second-order relativistic corrections 
in relative momenta $p$, $q$, which are determined in our approach by 
parameters $\omega_{1}$, $\tilde\omega_{1}$. In the case of $S$-states, the parameters 
$\omega_{n}$ are expressed through the momentum integrals $I_{n}$ in the form:
\begin{equation}
\label{f15}
I_{n}=\sqrt{\frac{2}{\pi}} \int_0^\infty p^2R(p)\frac{(\epsilon(p)+m)}{2\epsilon(p)}
\left(\frac{\epsilon(p)-m}{\epsilon(p)+m}\right)^n dp,~~~
\omega_{1}=\frac{I_{1}}{I_{0}},~\omega_{2}=\frac{I_{2}}{I_{0}},
\end{equation}
where $R(p)$ is radial wave function of charmonium,
\begin{equation}
I_{0} = \tilde R(0)=\sqrt{\frac{2}{\pi}}\int_0^\infty \frac{(\epsilon(p)+m)}
{2\epsilon(p)}p^2R(p)dp.
\end{equation}

In the case of $P$-states, relativistic parameters $\tilde\omega_{n}$ are determined 
by the momentum integrals $J_{n}$:
\begin{equation}
\label{f16}
J_{n}=\frac{\sqrt{2}}{3\sqrt{\pi}}\int_0^\infty q^3R(q)\frac{(\epsilon(q)+m)}{2\epsilon_(q)}
\left(\frac{\epsilon(q)-m}{\epsilon(q)+m}\right)^n dq,~~~
\tilde\omega_{1}=\frac{J_{1}}{J_{0}},~~~\tilde\omega_{2}=\frac{J_{2}}{J_{0}},
\end{equation}
\begin{equation}
J_{0} = \tilde R'(0)=\frac{\sqrt{2}}{3\sqrt{\pi}}\int_0^\infty \frac{(\epsilon(q)+m)}
{2\epsilon(q)}q^3 R(q)dq.
\end{equation}

We neglect higher order relativistic effects that are determined by parameters $\omega_n$, $\tilde\omega_n$
$(n\ge 2)$.
Numerical values of relativistic parameters $\omega_{1}$ and $\tilde{\omega}_{1}$, 
as well as relativistic wave functions at zero $\tilde{R}(0)$ and $\tilde{R}'(0)$ 
are given in Table~\ref{tab_mes_const}.
\begin{table}[htbp]
\caption{Basic parameters of the $S$ - and $P$ - states of charmonium.}
\bigskip
\label{tab_mes_const}
\begin{tabular}{|c|c|c|c|c|c|c|}
\hline
Meson            & $J^{PC}$   & Mass, MeV  & $\omega_1$ & $\tilde{\omega}_1$ & 
$\tilde{R}(0) \text{ GeV}^{3/2}$ & $\tilde{R}'(0) \text{ GeV}^{5/2}$ \\ 
\hline
$J/\Psi(1S)$     & $1^{--}$   & $3096.900$ & $0.20$ & ---    & $0.81$ & ---      \\  
$\eta_c(1S)$     & $0^{-+}$   & $2984.1$   & $0.20$ & ---    & $0.92$ & ---      \\  
$\chi_{c0}(1P)$  & $0^{++}$   & $3414.71$  & ---    & $0.04$ & ---    & $0.33$   \\  
$\chi_{c1}(1P)$  & $1^{++}$   & $3510.67$  & ---    & $0.05$ & ---    & $0.20$   \\  
$\chi_{c2}(1P)$  & $2^{++}$   & $3556.17$  & ---    & $0.07$ & ---    & $0.13$   \\  
$h_c(1P)$        & $1^{+-}$   & $2525.37$  & ---    & $0.06$ & ---    & $0.17$   \\  
\hline
\end{tabular}
\end{table}

The differential width of the Higgs boson decay into a pair of charmoniums takes the form:
\begin{equation}
\label{f16a}
d\Gamma=\frac{|{\bf P}|}{32\pi^2 M_H^2} \overline{|{\cal M}(H\to M_1 + M_2)|^2}d\Omega,
\end{equation}
where the modulus of the charmonium momentum vector is expressed in terms of the masses of the S- and 
P-states and the mass of the Higgs boson:
\begin{equation}
\label{f16b}
|{\bf P}|=\frac{1}{2M_H}\sqrt{[M_H^2-(M_1-M_2)^2][M_H^2-(M_1+M_2)^2]}.
\end{equation}

Using the amplitudes \eqref{f10} and \eqref{f13}, as well as the expression for the width \eqref{f16a}, 
we obtain total decay widths of the processes $H\to ZZ\to \eta_c+\chi_{c1}$ 
and $H\to ZZ\to J/\Psi+h_{c}$ in the following $\omega_n$form:
\begin{equation}
\label{f16c}
\Gamma(H\to ZZ\to \eta_c+\chi_{c1})=\frac{6 \alpha^3 |{\bf P}|
[1-(r_1+r_2)^2]}{ m^2 M_Z^6\sin^6 2\theta_W(1-r_4^2)^2}
(1-\omega_{10}-\omega_{01}+\omega_{10}\omega_{01})^2 |\tilde R'_P(0)|^2|\tilde R_S(0)|^2\times
\end{equation}
\begin{displaymath}
\Bigl[1+\frac{(1-r_4^2)(-\frac{1}{2}+a_c-a_c^2)}{r_1r_2(4-r_5^2)^2}
\bigl[(r_1^2+r_2^2-1)r_3r_4(r_3^2-r_4^2)-r_1r_2(r_3^4-8r_3^2-r_4^4+8r_4^2-32)\bigr]\Bigr]^2,
\end{displaymath}
\begin{equation}
\label{f16d}
\Gamma(H\to ZZ\to J/\Psi+h_{c})=\frac{6\alpha^3 |{\bf P}|
[1-(r_1+r_2)^2]a^2_z(1-a_c)^2}{m^2 M_Z^6\sin^6 2\theta_W(4-r_5^2)^4 r_1^2r_2^2}
(1-\omega_{10}-\omega_{01}+\omega_{10}\omega_{01})^2 \times
\end{equation}
\begin{displaymath}
|\tilde R'_P(0)|^2|\tilde R_S(0)|^2(r_4^2-r_3^2)^2
\bigl[r_3r_4(1-r_1^2-r_2^2)+r_1r_2(r_3^2+r_4^2-8)\bigr]^2.
\end{displaymath}
Numerical values of relative decay widths in non-relativistic approximation and 
taking into account relativistic corrections are presented in Table~\ref{tb1}.

Relativistic corrections in this mechanism factorize into decay widths in the form
$(1-\omega_{10}-\omega_{01}+\omega_{10}\omega_{01})^2$, which leads to a decrease in nonrelativistic 
decay amplitude by approximately two times for charmonium states.
The decay width $H\to ZZ\to J/\Psi+h_{c}$ (see table \ref{tb1}) is suppressed compared to the decay 
$H\to ZZ\to \eta_c+\chi_{c1}$ by several orders of magnitude. This is due to both the vanishing 
of direct decay amplitude and the appearance of additional small factors 
$a_c$, $(r_4^2-r_3^2)$ in it.
\begin{table}[htbp]
\caption{Numerical results for relative decay widths in non-relativistic approximation 
and taking into account relativistic corrections.}
\bigskip
\label{tb1}
\begin{tabular}{|c|c|c|}
\hline
Decay                         & ${\cal B}r_{nr}$    & ${\cal B}r_{rel}$  \\  \hline
$H\to ZZ\to \eta_c+\chi_{c1}$              & $0.62\cdot 10^{-14}$    &  $0.19\cdot 10^{-14}$  \\  
$H\to ZZ\to J/\Psi+h_{c}$                  & $0.95\cdot 10^{-22}$    &  $0.10\cdot 10^{-22}$    \\  
$H\to c{\bar c} \gamma \to J/\Psi+h_{c}$   & $1.34\cdot 10^{-11}$  &  $0.26\cdot 10^{-11}$  \\ 
$H\to c{\bar c} g \to J/\Psi+h_{c}$        & $1.04\cdot 10^{-14}$   &  $0.20\cdot 10^{-14}$ \\  
$H\to c{\bar c} g \to \eta_c + \chi_{c1}$  & $0.89\cdot 10^{-14}$ &  $1.48\cdot 10^{-14}$ \\  
$H\to ZZ\to T_{cc\bar c\bar c}+\gamma$      & $0.38\cdot 10^{-14}$ &  $0.21\cdot 10^{-14}$ \\   \hline
\end{tabular}
\end{table}

In the study of decay processes of the Higgs boson into a pair of $S$ - charmonium states 
\cite{faustov2023relativistic} it was shown that along with the $ZZ$-mechanism, the quark-photon 
and quark-gluon decay mechanisms are important, the amplitudes of which are shown in Fig.~\ref{qph_mech} 
and \ref{qgl_mech}. In the case of the quark-photon mechanism, the laws of conservation of angular momentum, 
spatial and charge parity limit the possible pairs of final states to only one: $H\rightarrow J/\Psi+h_c$. 
In this case, $J/\Psi$ is directly produced from a photon. The amplitude of quark-photon mechanism 
of pair production is obtained in the form:
\begin{equation}
\label{f16f}
{\cal M}(H\to c{\bar c} \gamma \rightarrow J/\Psi + h_c) = 
\frac{32 \cdot 2^{1/4} \alpha \sqrt{\pi r_2 G_F}\tilde R'_P(0) \tilde R_S(0)}
{\sqrt{3 r_1} M_H^2} \times
\end{equation}
\begin{equation*}
(3+\omega_1)(-1+\tilde{\omega_1})\varepsilon_{\mu\nu\lambda\sigma}
v_1^\mu v_2^\nu \varepsilon^\lambda(v_2,J_z)\varepsilon^\sigma(v_1,S_z),
\end{equation*}
where $G_F$ is the Fermi constant: $(\sqrt{2}G_F)^{1/2}=e/2M_W\sin\theta_W$,
$\varepsilon(v_2,J_z)$ is the polarization vector of the $h_c$ meson,
$\varepsilon(v_1,S_z)$ is the polarization vector of the $J/\Psi$ meson,
$M_1,~M_2$ are the masses of the $J/\Psi$ and $h_c$ mesons, respectively.
Calculating the square of the amplitude modulus further, we obtain total 
width for this process:
\begin{equation}
\label{f16h}
\Gamma(H\to c{\bar c} \gamma \rightarrow J/\Psi + h_c) = 
\frac{64 \sqrt{2} \alpha^2 G_F
|{\bf P}| |\tilde R'_P(0)|^2 |\tilde R_S(0)|^2}{3 M_H^6 r_1^3 r_2} \times
\end{equation}
\begin{equation*}
(r_1^4+(r_2^2-1)^2-2r_1^2(1+r_2^2)) (3+\omega_1)^2(1-\tilde{\omega_1})^2.
\end{equation*}

The second quark - gluon mechanism is shown in Fig.~\ref{qgl_mech}. In this case, 
two different combinations of the pair of final states of mesons are possible: 
$ H \rightarrow J/\Psi + h_c $, $ H \rightarrow \eta_c + \chi_{c1}$. 
The amplitudes corresponding to these processes are obtained in our approach in the form:
\begin{figure}[htbp]
\centering
\includegraphics[width=0.6\textwidth]{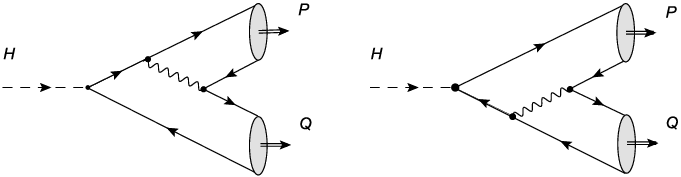}
\caption{Quark - gluon mechanism of charmonium pair production. $P$, $Q$ are four-momenta 
of quarkonia.}
\label{qgl_mech}
\end{figure}
\begin{equation}
\label{f16i}
{\cal M}(H\to c{\bar c} g \rightarrow J/\Psi + h_c) = \frac{256 \cdot 2^{1/4} \alpha_s r_1 
\sqrt{\pi r_1 r_2 G_F } \tilde{R}'_P(0) \tilde{R}_S(0) }{3\sqrt{3} M_H^2} \times
\end{equation}
\begin{equation*}
(3+\omega_1)(1-\tilde{\omega_1})
\varepsilon_{\mu\nu\lambda\sigma}v_1^\mu v_2^\nu \varepsilon^\lambda(v_2,J_z)\varepsilon^\sigma(v_1,S_z),
\end{equation*}
\begin{equation}
\label{f16j}
{\cal M}(H\to c{\bar c} g \rightarrow \eta_c + \chi_{c1}) = \frac{128 \cdot 2^{3/4} 
\alpha_s \sqrt{\pi r_1 r_2 G_F } \tilde{R}'_P(0) \tilde{R}_S(0) }{\sqrt{3} M_H^2} \times 
\end{equation}
\begin{equation*}
\left( 2 r_6(-3 +7 \omega_1) +5  r_1 (1 +\omega_1) \right)(-1+\tilde{\omega}_1) 
\left( v_1 \varepsilon(v_2,J_z) \right),~~~r_6=m/M_H.
\end{equation*}

The value of the constant $\alpha_s(\frac{1}{2}M_H) = 0.1245$ is taken into account with 
the two-loop contribution \cite{kniehl}. 
Total widths corresponding to these amplitudes take the form:
\begin{equation}
\label{f16l}
\Gamma(H\to c{\bar c} g \rightarrow J/\Psi + h_c) = 
\frac{4096 \sqrt{2} \alpha_s^2 G_F r_1 |{\bf P}| |\tilde R'_P(0)|^2 |\tilde R_S(0)|^2}
{27 M_H^6 r_2} \times
\end{equation}
\begin{equation*}
(r_1^4+(r_2^2-1)^2-2r_1^2(1+r_2^2))(3+\omega_1)^2(-1+\tilde{\omega_1})^2,
\end{equation*}
\begin{equation}
\label{f16m}
\Gamma(H\to c{\bar c} g \rightarrow \eta_c + \chi_{c1}) = 
\frac{1024 \sqrt{2} \alpha_s^2 G_F |{\bf P}| |\tilde R'_P(0)|^2 |\tilde R_S(0)|^2}{3 M_H^6 r_1 r_2} \times
\end{equation}
\begin{equation*}
(r_1^4+(r_2^2-1)^2-2r_1^2(1+r_2^2)) \left( 2r_6 (-3 +7 \omega_1) +5 r_1 (1 + \omega_1) 
\right)^2(-1+\tilde{\omega_1})^2.
\end{equation*}
Numerical values of decay widths \eqref{f16h}, \eqref{f16l} and \eqref{f16m} 
in non-relativistic approximation and taking into account relativistic corrections 
are included in Table~\ref{tb1}.

\section{Production of fully charmed tetraquark}

The approach considered for the production of heavy quark bound states can be used to describe the production 
of heavy baryons and tetraquarks. We consider next the production of a vector tetraquark consisting 
of two $c$ quarks and two $\bar c$ antiquarks in the decay of 
$H \to T_{cc\bar c\bar c}+\gamma$. The LHCb collaboration detected the resonance $X(6900)$, which was confirmed 
by the ATLAS experiment and is a candidate for the tetraquark $(cc\bar c\bar c)$ along with numerous 
other resonances discovered in recent years \cite{t1,t2,t3,t3a} (see other references in \cite{t3,t3a}).

A discovery of charmonium-like state $X(3872)$ by the Belle collaboration \cite{t1} in 2003 brought the study of 
exotic states to the forefront of hadron physics. Although many tetraquark 
candidates containing c, b quarks are now known, many questions remain about 
them, including which ones are truly exotic hadrons and what their internal structure is
\cite{t4,t5,t14,t6,t7}.

In connection with the emerging experimental data on the production of exotic bound states 
of quarks and gluons, a significant number of papers have appeared in which the cross sections 
for the production of heavy tetraquarks in proton-proton interactions, in electron-positron annihilation, 
and in B-meson decays \cite{t8,t9,t10,t11,t12,t13,t14a,t14b} have been calculated. 
We have investigated the production of a tetraquark in the decay of the Higgs boson 
$H\to T_{cc\bar c\bar c}+\gamma$ using the $ZZ$-boson production mechanism 
(see the right amplitude in Fig.~\ref{zz_mech}).

In the case where a tetraquark consists of identical quarks and antiquarks, the Pauli principle leads 
to additional restrictions on the quantum numbers of a pair of identical quarks (a diquark). 
Indeed, permutation of identical quarks should change the sign of the total wave function of quarks.
We consider the antisymmetric color part of the wave function of a pair of identical quarks and antiquarks. 
The coordinate wave function 
is symmetric because the quarks are in the $S$-wave state, so the spin part of the wave function 
should also be symmetric. Consequently, total spin of the $S$-wave state of a pair of quarks should 
be equal to 1.

There are various possible configurations of four quarks and antiquarks that lead to the formation of a tetraquark. 
We consider the case of production when a tetraquark is formed from two pairs, 
quark-quark and antiquark-antiquark, each of which has spin 1 and is in a color-antisymmetric state. 
Introducing total $P$, $Q$ and relative $p$, $q$ momenta in these pairs, 
we can express the momenta of individual quarks and antiquarks using \eqref{f6a}. 
We further introduce total $T$ and relative $t$ momenta for these pairs:
\begin{equation}
\label{f18}
P=\frac{1}{2}T+t,~~~Q=\frac{1}{2}T-t.
\end{equation}

Denoting the photon momentum as $k$ and the Higgs boson momentum as $r$, we obtain:
\begin{equation}
\label{f19}
r=k+T,~~~kT=\frac{1}{2}(M_H^2-M_T^2),~~~rT=\frac{1}{2}(M_H^2+M_T^2).
\end{equation}

The spin of a tetraquark, obtained by adding two spins of 1 pairs $(cc)$ and $(\bar c\bar c)$, 
can take three values J=0, 1, 2. The photon is emitted by one of the quarks or antiquarks, 
so we have four decay amplitudes, one of which is shown in right Fig.~\ref{zz_mech}. 
General initial expressions for these amplitudes are:
\begin{equation}
\label{f20}
{\cal M}_{T+\gamma}^{(1)}=\frac{e^4}{4\sin 2\theta M_Z m^2}\int d{\bf p}\int d{\bf q}\int d{\bf t}
\Psi_T({\bf p,\bf q,\bf t})D^{\alpha\lambda}_Z(k_1)D^{\alpha\sigma}_Z(k_2)\times
\end{equation}
\begin{displaymath}
\frac{N({\bf p,\bf t})N({\bf q,\bf t})}{\bigl[\frac{1}{4}\bigl(M_H^2-
\frac{3}{4}M_T^2\bigr)-m^2\bigr]}
Tr \Bigl\{
\hat\Pi_1^V({\bf p,\bf t})\hat\varepsilon_\gamma(\hat p_1+\hat k+m)\Gamma^\lambda
\hat\Pi_2^V({\bf q,\bf t})\Gamma^\sigma \Bigr\},
\end{displaymath}
\begin{equation}
\label{f21}
{\cal M}_{T+\gamma}^{(2)}=\frac{e^4}{4\sin 2\theta M_Z m^2}\int d{\bf p}\int d{\bf q}\int d{\bf t}
\Psi_T({\bf p,\bf q,\bf t})D^{\alpha\lambda}_Z(k_1)D^{\alpha\sigma}_Z(k_2)\times
\end{equation}
\begin{displaymath}
\frac{N({\bf p,\bf t})N({\bf q,\bf t})}{\bigl[\frac{1}{4}\bigl(M_H^2-
\frac{3}{4}M_T^2\bigr)-m^2\bigr]}
Tr \Bigl\{
\hat\Pi_1^V({\bf p,\bf t})\Gamma^\lambda
\hat\Pi_2^V({\bf q,\bf t})\Gamma^\sigma (\hat p_1+\hat k+m)\hat\varepsilon_\gamma \Bigr\},
\end{displaymath}
\begin{equation}
\label{f22}
{\cal M}_{T+\gamma}^{(3)}=\frac{e^4}{4\sin 2\theta M_Z m^2}\int d{\bf p}\int d{\bf q}\int d{\bf t}
\Psi_T({\bf p,\bf q,\bf t})D^{\alpha\lambda}_Z(k_1)D^{\alpha\sigma}_Z(k_2)\times
\end{equation}
\begin{displaymath}
\frac{N({\bf p,\bf t})N({\bf q,\bf t})}{\bigl[\frac{1}{4}\bigl(M_H^2-
\frac{3}{4}M_T^2\bigr)-m^2\bigr]}
Tr \Bigl\{
\hat\Pi_1^V({\bf p,\bf t})\Gamma^\lambda (-\hat q_1-\hat k+m)\hat\varepsilon_\gamma
\hat\Pi_2^V({\bf q,\bf t})\Gamma^\sigma  \Bigr\},
\end{displaymath}
\begin{equation}
\label{f23}
{\cal M}_{T+\gamma}^{(4)}=\frac{e^4}{4\sin 2\theta M_Z m^2}\int d{\bf p}\int d{\bf q}\int 
d{\bf t} \Psi_T({\bf p,\bf q,\bf t})D^{\alpha\lambda}_Z(k_1)D^{\alpha\sigma}_Z(k_2)\times
\end{equation}
\begin{displaymath}
\frac{N({\bf p,\bf t})N({\bf q,\bf t})}{\bigl[\frac{1}{4}\bigl(M_H^2-
\frac{3}{4}M_T^2\bigr)-m^2\bigr]}
Tr \Bigl\{
\hat\Pi_1^V({\bf p,\bf t})\Gamma^\lambda 
\hat\Pi_2^V({\bf q,\bf t})\hat\varepsilon_\gamma(-\hat q_2-\hat k+m)
\Gamma^\sigma  \Bigr\},
\end{displaymath}
where relativistic normalization factor is equal to
\begin{equation}
\label{f24}
N({\bf p,\bf t})=\left[
\frac{\epsilon({\bf p}+{\bf t/2})}{m}\frac{(\epsilon({\bf p}+{\bf t/2})+m)}{2m}
\frac{\epsilon({\bf p}-{\bf t/2})}{m}\frac{(\epsilon({\bf p}-{\bf t/2})+m)}{2m}
\right]^{-1/2},
\end{equation}
and $\Psi_T({\bf p,\bf q,\bf t})$ is the wave function of the tetraquark in the rest frame 
in momentum representation. The momenta in $Z$-boson propagators have the form:
\begin{equation}
\label{f25}
k_1=\frac{1}{2}T+p+q+t,~~~k_2=\frac{1}{2}T-p-q.
\end{equation}

The amplitudes \eqref{f20}-\eqref{f23} contain projection operators onto the state of a pair 
of quarks or a pair of antiquarks with spin 1:
\begin{equation}
\label{f26}
\hat\Pi_i^V=
\bigl[\frac{\hat v-1}{2}-\hat v\frac{(\epsilon({\bf p}+\frac{\bf t}{2})-m)}{2m}-
\frac{(\hat p+\frac{\hat t}{2})}{2m}\bigr]\hat{\varepsilon_i}(1+\hat v) 
\bigl[\frac{\hat v+1}{2}-\hat v\frac{(\epsilon({\bf p}-\frac{\bf t}{2})-m)}{2m}+
\frac{(\hat p-\frac{\hat t}{2})}{2m}\bigr],
\end{equation}
where $\varepsilon_i$ (i=1,2) are the polarization vectors describing pairs of quarks and antiquarks 
with spin 1. The color part of the amplitudes \eqref{f20}-\eqref{f23} is 
distinguished as a product of color factors describing pairs of quarks and antiquarks 
in an antisymmetric state.

Keeping in the numerator of amplitudes the quadratic terms in relative momenta 
$\omega_p$ = $<\frac{{\bf p}^2}{m^2}>$,
$\omega_q=<\frac{{\bf q}^2}{m^2}>$, $\omega_t=<\frac{{\bf t}^2}{m^2}>$,
we obtain following expression for the numerator of total amplitude of the process 
of a vector tetraquark and photon production:
\begin{equation}
\label{f27}
{\cal N}_{T+\gamma}=N_1(v\varepsilon_\gamma)(v_\gamma\varepsilon_T)+
N_2(\varepsilon_\gamma\varepsilon_T)+N_3\varepsilon_{\mu\nu\sigma\lambda}
v^\mu v_\gamma^\nu \varepsilon_T^\sigma \varepsilon_\gamma^\lambda,
\end{equation}
\begin{equation}
\label{f28}
N_1=(1-2a_c)\bigl(1-\frac{7}{48}r_2^2+\frac{7}{48}r_3^2-\frac{1}{192}r_2^2 
r_3^2\bigr)\omega_t,
\end{equation}
\begin{equation}
\label{f29}
N_2=(1-2a_c)\omega_t\Bigl[r_3^{-1}\bigl(-\frac{1}{2}\frac{r_2^2}{r_3}-
\frac{1}{96}\frac{r_2^4}{r_3}\bigr)+\frac{1}{4}-
\frac{1}{24}r_2^2+\frac{1}{128}r_2^4+\frac{1}{32}r_3^2-\frac{1}{192}r_3^2r_2^2\Bigr],
\end{equation}
\begin{equation}
\label{f30N}
N_3=32-2r_2^2-4r_3^2+\frac{1}{2}r_2^2r_3^2-a_c(1-a_c)+(\omega_p+\omega_q)\Bigl[
-\frac{112}{3}+\frac{7}{3}r_2^2+\frac{14}{3}r_3^2-\frac{7}{12}r_2^2r_3^2+
\end{equation}
\begin{displaymath}
a_c(1-a_c)\bigl(\frac{224}{3}+\frac{2}{3}r_2^2+\frac{4}{3}r_3^2-\frac{1}{6}r_2^2r_3^2\bigr)\Bigr]+
\omega_t\Bigl[-\frac{61}{3}+\frac{59}{48}r_2^2+\frac{67}{24}r_3^2-\frac{65}{192}r_2^2 r_3^2+
\end{displaymath}
\begin{displaymath}
a_c(1-a_c)\bigl(\frac{122}{3}-\frac{1}{2}r_2^2-r_3^2+\frac{1}{8}r_2^2 r_3^2\bigr)
\Bigr],
\end{displaymath}
where the auxiliary four-vectors are introduced: $v=T/M_T$, $v_\gamma=k/M_T$, 
$\varepsilon_T$ is the tetraquark polarization vector. The decay width 
taking into account the leading-order relativistic corrections is equal to:
\begin{equation}
\label{f30}
\Gamma(H\to ZZ\to \gamma+T)=
\frac{256\pi^3\alpha^4 r_3^2(r_2^2-r_3^2)|\Psi(0,0,0)|^2}{\sin^2 2\theta_W M_Z^8 r_2^3(4-r_3^2)^2
(2r_2^2-r_3^2-4)^2[\frac{1}{4}(r_2^2-\frac{3}{4}r_3^2)-r_1^2r_3^2]^2}\times
\end{equation}
\begin{displaymath}
\left[3N_2^2+2N_3^2(vv_\gamma)^2-N_1^2(vv_\gamma)^2\right],~~~(vv_\gamma)=
(r_2^2-r_3^2)/2r_3^2.
\end{displaymath}

The wave function of the tetraquark was obtained within the variational method 
as in the case of three particles \cite{t15}
in the form ($\boldsymbol\rho,\boldsymbol\lambda,\boldsymbol\sigma $ 
are the Jacobi coordinates):
\begin{equation}
\label{f31}
\Psi(\boldsymbol\rho,\boldsymbol\lambda,\boldsymbol\sigma)=\sum_{I=1}^K
C_I e^{-\frac{1}{2}\bigl[A_{11}(I)
\boldsymbol\rho^2+
2A_{12}(I)\boldsymbol\rho\boldsymbol\lambda+A_{22}(I)\boldsymbol\lambda^2+
2A_{13}(I)\boldsymbol\rho\boldsymbol\sigma+2A_{23}(I)\boldsymbol\lambda\boldsymbol\sigma+
A_{33}(I)\boldsymbol\sigma^2\bigr]},
\end{equation}
where $A_{ij}(I)$ is the matrix of nonlinear variational parameters, $C_I$ are linear 
variational parameters. In momentum representation, the wave function of the system 
has the form:
\begin{equation}
\label{f33}
\Psi({\bf p},{\bf q},{\bf t})=\sum_{I=1}^K\frac{C_I}{\sqrt{{\cal N}}(det A)^{3/2}(2\pi)^{\frac{9}{2}}}
e^{-\frac{1}{2det A}[{\bf p}^2(A_{22}A_{33}-
A_{23}^2)+{\bf q}^2(A_{11}A_{33}-A_{13}^2)+{\bf t}^2(A_{11}A_{22}-A_{12}^2)]}
\end{equation}
\begin{displaymath}
e^{-\frac{1}{2det A}[2{\bf p}{\bf t}(A_{12}A_{23}-A_{13}A_{22})+
2{\bf q}{\bf t}(A_{12}A_{13}-A_{11}A_{23})+2{\bf p}{\bf q}(A_{13}A_{23}-A_{12}A_{33})]},
\end{displaymath}
${\bf p}$, ${\bf q}$, ${\bf t}$ are relative momenta for particles 12, 34 and the pair of particles 12 and 34, 
matrix elements $A_{kl}=A_{kl}(I)$. Using \eqref{f33}, the values 
of parameters in \eqref{f30} are obtained:
\begin{equation}
\label{f40}
\Psi(0,0,0)=0.10~GeV^{9/2}, ~\omega_p=\langle\frac{{\bf p}^2}{m^2}\rangle=0.18,~
\omega_q=\langle\frac{{\bf q}^2}{m^2}\rangle=0.18,~
\omega_t=\langle\frac{{\bf t}^2}{m^2}\rangle=0.12.
\end{equation}
A more detailed calculation of the wave functions and parameters \eqref{f40} is presented in \cite{tetra2025}.
Other relativistic corrections, which are determined by $\langle\frac{{\bf p}{\bf t}}{m^2}\rangle$,
$\langle\frac{{\bf q}{\bf t}}{m^2}\rangle$, $\langle\frac{{\bf p}{\bf q}}{m^2}\rangle$, 
are not given here, since their value turns out to be two orders of magnitude smaller 
than those considered \eqref{f40}.
Numerical value of decay width $H\to T_{cc\bar c\bar c}+\gamma$ is presented 
in Table~\ref{tb1}.

\section{Conclusion}

Theoretical investigation of rare exclusive decays of the Higgs boson 
with the production of quark bound states is an important problem 
in studying the Higgs sector in the Standard Model. 
Such processes are considered as one of the possibilities for searching for new physics beyond 
the Standard Model. In addition, such decays may help to improve our understanding of quark-gluon 
dynamics at short distances. In this work, our investigation of processes 
of pair charmonium production in H-boson decays \cite{faustov2023relativistic} 
is extended to the case of S- and P-wave charmonium states. 
We also use relativistic method for describing the production of quark bound states
to calculate the decay width of the H-boson with the formation of the tetraquark 
$(cc\bar c\bar c)$. 

The obtained expressions for the decay amplitudes and widths of the H-boson 
contain the mass and relativistic parameters calculated within the framework of relativistic quark model. 
The results in Table~\ref{tb1} show that the largest decay width is obtained for the 
process $H\to c{\bar c} \gamma \to J/\Psi+h_{c}$. Relativistic corrections, which are determined 
using the parameters $\omega_n$, $\tilde\omega_n$, $\tilde R(0)$, $\tilde R'(0)$, significantly 
change the results of calculation in non-relativistic approximation. 
The systematic decrease of relativistic decay widths compared to nonrelativistic ones in pair production 
of charmonium is determined mainly by the value of the wave function of the bound states at zero. 
Relativistic corrections in the quark interaction potential lead to a decrease of this value, 
which enters the decay width to the fourth power.
Relativistic corrections must be taken 
into account in the case of c-quark production to obtain reliable predictions for the decay widths.
The obtained estimate for the Higgs boson decay width in the $H\to T(cc\bar c\bar c)+\gamma$ process 
of vector tetraquark production shows 
that the probability of such a reaction is comparable to the probability of the processes of pair 
production of charmonium.
Experimental observation of such rare decays may be possible in the future at high-luminosity colliders.
This estimate of the decay width was obtained as a result of calculating the tetraquark wave function 
within the framework of the variational method with a Gaussian basis \cite{t15}.
The calculation of the mass spectrum and wave functions of heavy tetraquarks will be presented in another paper.
It is necessary to study other decay mechanisms with the formation of tetraquarks, which is what 
is planned to be done in the future.

\begin{acknowledgments}
This work was supported by the Foundation for the Development of Theoretical Physics and Mathematics 
BASIS (grant 22-1-1-23-1).
\end{acknowledgments}

\end{document}